\documentclass[%
 reprint,
 amsmath,amssymb,
 aps,
 pra,
]{revtex4-2}

\usepackage{graphicx,lipsum}
\usepackage{dcolumn}
\usepackage{bm}
\usepackage{subfigure}
\usepackage{multirow}
\usepackage{makecell}
\usepackage[section]{placeins}
\usepackage{multirow}
\usepackage{braket}
\usepackage[margin=1.5cm]{geometry}
\usepackage[colorlinks=true,linkcolor=blue,citecolor=blue,urlcolor=blue,bookmarks=false,hypertexnames=true]{hyperref}
\usepackage{epsf,epsfig,graphics}
\usepackage{orcidlink}
\usepackage[utf8]{inputenc}
\begin{document}

\preprint{APS/123-QED}

\title{Interaction-rotation driven localization-delocalization crossover in Fock space: An exact diagonalization study on trapped Bose gas}

\author{Mohd Talib \orcidlink{0000-0001-7261-3584}}
\email{rs.mtalib@jmi.ac.in}
\author{M. A. H. Ahsan \orcidlink{0000-0002-9870-2769}}%
\email{mahsan@jmi.ac.in}
\affiliation{Department of Physics, Jamia Millia Islamia, New Delhi, 110025, India.}%

\date{\today}

\begin{abstract}
We study the localization-delocalization crossover and quantum entanglement with varying numbers of interacting bosons for the nonrotating and the rotating cases. The many-body ground state within subspaces of given total angular momentum is obtained via exact diagonalization and analyzed using the inverse participation ratio (IPR), the information (Shannon) entropy and the von Neumann entanglement entropy. In the non-rotating case, a crossover from localization to delocalization is observed with increasing interaction, characterized by decreasing value of IPR with corresponding increase in information entropy and von Neumann entanglement entropy, arising from interaction-induced depletion of Bose-Einstein condensate.  When subjected to rotation, the system is driven further towards delocalization due to rotation-induced depletion of Bose-Einstein condensate. The crossover from localization to delocalization is indicated by distribution of the eigenstate weights over an increasing number of basis states in the many-body Hilbert space. The consistent behavior of IPR, information entropy and von Neumann entanglement entropy with respect to interaction and rotation demonstrates that these quantities provide a unified characterization of the localization-delocalization crossover. Results on computed von Neumann entropy show that localized states exhibit weak entanglement while delocalized states are characterized by strong entanglement within the quantum many-body system.  
\end{abstract}

\maketitle

\section{Introduction} 
Anderson localization, introduced to describe single-particle localization in a non-interacting system in presence of disorder \cite{Anderson1958}, has since been extended to interacting many-body systems, giving rise to the phenomena of many-body localization (MBL) \cite{Sierant2025, Roy2025, Abanin2019, Chen2023, dupont2026} for a large class of systems, including Heisenberg spins \cite{David2015}, Bose-Hubbard model \cite{Hopjan2020, Sierant2017}, and Floquet systems \cite{Sierant2023, Sonner2021}. The MBL provides an instance for the breakdown of the eigenstate thermalization hypothesis (ETH), according to which non-integrable quantum many-body systems evolve towards thermal equilibrium under their own unitary dynamics \cite{Deutsch1991, Srednicki1994, Polkovnikov2011}. 
There have as well been numerous experimental studies to explore a wide range of physical phenomena such as the observation of MBL in linear chains of nuclear spins in fluorapatite $[\mathrm{Ca_{5}(PO_{4})_{3}F}]$ where each $\mathrm{F}$ spin is surrounded by three $\mathrm{P}$ spins \cite{Ken2018}, to investigate the role of quantum entanglement in facilitating the emergence of statistical mechanics in a pure quantum state \cite{Adam2016}, the universal signatures of many-body quantum chaos in quantum devices \cite{Hang2025}, etc. 

With experimental advances in cooling and trapping \cite{Phillips1998}, clouds of atomic vapours with tunable interatomic interaction via Feshbach resonance \cite{inouye1998, chin2010}, it has become possible to generate quantized vortices using focused laser beams \cite{Madison}. This leads us to consider a system of trapped interacting bosons to examine the effects of interaction and rotation on many-body localization-delocalization crossover and quantum entanglement within the Bose system which has remained largely unexplored. To this end, we employ three measures---the inverse participation ratio (IPR) \cite{Misguich2016, Emmanuel2017, Liu2025, Stephan2009, Stephan2011}, the information (Shannon) entropy \cite{Lea2010shannon}, and the von Neumann entanglement entropy \cite{Islam2015}---to study the many-body localization-delocalization crossover and entanglement within the many-body ground state. The IPR has been employed as an indicator of ergodicity breaking in a model exhibiting MBL \cite{Luca_2013}, as a criterion for identifying Anderson localization in disordered systems \cite{Edwards_1972}, to reveal scaling behavior across integrable-nonintegrable regimes in XXZ spin chain \cite{Misguich2016}, as a tool to relate the XX Heisenberg spin chain with circular symplectic ensemble of Dyson and the Dyson-Gaudin Coulomb lattice gas \cite{Emmanuel2017}. The information entropy provides a means to investigate the breaking of integrability leading to transition from superfluid to insulator in the ground state of hardcore bosons \cite{Lea2010shannon}, a measure of delocalization and the consequent viability of thermalization in bosonic as well as fermionic systems \cite{Santos2010thermal}, to examine the effect of the local and global quenches in driving a system of interacting spins into the chaotic regime leading to exponential fidelity decay indicating the onset of thermalization \cite{Torres2014}. Further, the entanglement entropy provides an insight into a variety of phenomena, including several of the quantum phase transitions such as metal-insulator transition \cite{Pouranvari2014}, many-body localization transition in the presence of disorder \cite{Geraedts_2017}, measurement-induced transitions that provides a theoretical framework connecting information extraction and purification dynamics with classical percolation in the large-$q$ limit \cite{Yimu2020}, random projective measurements driving entangling-disentangling transition \cite{Brian2019}, and in the construction of the spatial geometry and curvature of quantum states \cite{ChunJun2017, RAAMSDONK2010}.  

The rest of the paper is organized as follows. Sec. \ref{1} introduces the model Hamiltonian for trapped interacting bosons. In Sec. \ref{2}, we discuss the physical quantities employed, specifically the inverse participation ratio, the information entropy and the von Neumann entanglement entropy. The numerical results are presented in Sec. \ref{3}, and Sec. \ref{4} concludes the paper with a summary. 

\section{The Model system} \label{1}
For a system of \(N\) interacting bosons confined harmonically in the \(\mathrm{xy}\)-plane, the effective Hamiltonian and the total angular momentum operator in the laboratory frame, can be expressed in dimensionless form as
\begin{align}\label{eq3}
\hat{H}^{\mathrm{lab}}= &\sum_{i=1}^{N}\left[\frac{1}{2}\left(\frac{a_{\perp}\nabla_{\perp i}}{i}\right)^{2} 
+ \frac{1}{2}\left(\frac{r_{\perp i}}{a_{\perp}}\right)^2 \right] \nonumber \\
&+
g_{2}\left( \frac{a_{\perp}}{\sqrt{2\pi}\sigma}\right)^{2}\sum_{i < j}^{N}\exp\!\left[-\frac{(r_{\perp i}-r_{\perp j})^{2}}{2\sigma^{2}}\right],
\end{align}
and
\begin{align*}
\hat{L}_{z}^{\mathrm{lab}}
= \frac{1}{i}\sum_{i=1}^{N}
\left(\mathbf{r}_{i}\times\nabla_{i}\right)_{z},
\end{align*}
respectively. Here, \(a_{\perp} \equiv \sqrt{\hbar/(M\omega_{\perp})}\) denotes the harmonic oscillator length associated with the radial trapping frequency \(\omega_{\perp}\), and \(M\) is the mass of the boson. In this work, all lengths are expressed in the units of trap length \(a_{\perp}\). In Eq. (\ref{eq3}), the first and the second terms represent the kinetic energy and the trap energy, respectively. The third term corresponds to the two-body interaction potential in \(\mathrm{xy}\)-plane, modeled by a finite-range Gaussian of width \(\sigma\) pre-multiplied by the dimensionless effective interaction strength defined as 
\begin{equation} \label{g2}
     g_{2} \equiv \frac{4\pi a_{sc}}{a_{\perp}} \sqrt{\frac{\lambda_{z}}{2\pi}},
\end{equation} 
where $a_{sc}$ is the $s$-wave scattering length and $\lambda_{z} \equiv \omega_{z}/\omega_{\perp}$ ($\gg$ 1, for a quasi-2D system) is the trap  anisotropy parameter, with $\omega_{z}$ denoting the axial trapping frequency. Throughout this work, we take $\lambda_{z}=4$. In the limit where the interaction range approaches zero i.e $\sigma \to 0 $, the Gaussian interaction reduces to the $\delta$-function (contact) potential $V(r_{\perp i},r_{\perp j})  \to g_{2}\delta(r_{\perp i}-r_{\perp j})$ \cite{dalfovo1999}.

When the system is subjected to an externally imposed rotation about the \(z\)-axis with angular velocity \(\Omega\), the Hamiltonian in the co-rotating frame becomes \cite{Ahsan2001, talib2025, talib2025spectral}
\begin{equation*}
    \hat{H}^{\mathrm{rot}} = \hat{H}^{\mathrm{lab}} - \Omega \hat{L}_{z}^{\mathrm{lab}}.
\end{equation*}
Here, \(\Omega\), expressed in units of \(\omega_{\perp}\), acts as a Lagrange multiplier that constrains the system to a subspace of the Hilbert space with a given total angular momentum \( \hat{L}_{z}^{\mathrm{lab}} \).

\section{Measures of localization and entanglement}\label{2}

The localization-delocalization in quantum systems relies on identifying quantitative measures that distinguish between the two regimes. Among the most widely used diagnostics are the inverse participation ratio (IPR), the information (Shannon) entropy and the von Neumann entanglement entropy. These quantities provide complementary perspectives on the characteristics of quantum states.

\subsection{Inverse participation ratio}
The IPR \cite{Bertsch1999,VISSCHER1972477, Misguich2016, Emmanuel2017, Torres2015, Soumya2015} serves as a quantitative measure of the localization-delocalization of quantum many-body eigenstates. A value of IPR close to unity indicates that the eigenstate is highly localized in the Hilbert space, with significant weight concentrated on only a few basis states, whereas smaller values of IPR correspond to increasingly delocalized states spread over a larger part of the Hilbert space. For a many-body quantum state \(\ket{\psi}\) in a \(D\)-dimensional Hilbert space \cite{talib2025},
\[
\ket{\psi} = \sum_{\nu=1}^{D} C_{\nu}\ket{\phi_{\nu}},
\]
where \(\ket{\phi_{\nu}}\) denotes the basis states and \(C_{\nu}\) are the corresponding expansion coefficients, it is defined as:

\begin{equation}
\text{IPR} = \sum_{\nu=1}^{D} |C_\nu|^4.
\end{equation}
For a localized state, IPR $\approx$ 1 \cite{Soumya2015, Mukherjee_2015}, while for a completely delocalized state, IPR $\approx$ $\frac{1}{D}$. 

\subsection{Information (Shannon) entropy}
The information entropy quantifies the extent of delocalization of the quantum state in the many-body Hilbert space and provides a measure of quantum fluctuations in the state. Lower values of entropy indicate that the eigenstate is localized over a few basis states in the Hilbert space, whereas higher values signify a uniform distribution over a larger part of the Hilbert space, reflecting enhanced delocalization.

The R\'enyi entropy provides a broader information-theoretic framework that captures the degree of delocalization and complexity inherent in a quantum state. The R\'enyi entropy of order $n$ ($n \geq 0$, $n \neq 1$) is defined as:

\begin{equation}
S_n = \frac{1}{1-n} \ln\left(\sum_{\nu=1}^{D} |C_\nu|^{2n}\right),
\end{equation}
for $n \to 1$, the Rényi entropy reduces to the information entropy:

\begin{equation}
S_1 = \lim_{n \to 1} S_n = -\sum_{\nu=1}^{D} |C_\nu|^{2} \ln |C_\nu|^{2},
\label{s1}
\end{equation}
for $n=2$, the Rényi entropy reduces to

\begin{equation}
    S_2 = -\ln\left(\sum_{\nu=1}^{D} |C_\nu|^4\right)= -\ln(\text{IPR}).
    \label{s2}
\end{equation}
For localized states, $\text{IPR} \approx 1$, and hence $S_2 \approx 0$. The information (Shannon) entropy $S_1$ in Eq. (\ref{s1}) and the second-order R\'enyi entropy (IPR) in Eq. (\ref{s2}) are functions of normalized coefficients $\{C_\nu\}$ averaged over the Hilbert space and are hence intensive quantities \cite{Jeynes2026}.  

\subsection{One-particle reduced density matrix}
The many-body ground-state wavefunction
$\Psi_0(\mathbf{r}_1,\mathbf{r}_2,\ldots,\mathbf{r}_N)$
for a system of $N$ bosons in a given subspace of total angular momentum $L_z$, obtained through exact diagonalization is assumed to be normalized to 1. The one-particle reduced density
matrix $\rho(\mathbf{r},\mathbf{r}')$ is obtained by integrating out
the degrees of freedom of the remaining $(N-1)$ particles \cite{Ahsan2001} and is given by
\begin{equation*}
\begin{aligned}
\rho(\mathbf{r},\mathbf{r}')
&=
\int d\mathbf{r}_2\,\cdots d\mathbf{r}_N\,
\Psi_0(\mathbf{r},\mathbf{r}_2,\ldots,\mathbf{r}_N)
\\
&\qquad\times
\Psi_0^*(\mathbf{r}',\mathbf{r}_2,\ldots,\mathbf{r}_N)
\\[4pt]
&=
\sum_{\nu=1}^{D}\sum_{\nu'=1}^{D}
C_{\nu} C^*_{\nu'}
\int d\mathbf{r}_2\,\cdots d\mathbf{r}_N\,
\phi_{\nu}(\mathbf{r},\mathbf{r}_2,\ldots,\mathbf{r}_N)
\\
&\qquad\times
\phi_{\nu'}^*(\mathbf{r}',\mathbf{r}_2,\ldots,\mathbf{r}_N),
\end{aligned}
\end{equation*}
and can be written in terms of a complete
set of single-particle basis functions $u_{\bf n}(\mathbf{r})$ as

\begin{equation*}
\rho(\mathbf{r},\mathbf{r}')
=
\sum_{\bf n, \bf n'} \rho_{\bf n, \bf n'}\,
u_{\bf n}(\mathbf{r})u_{\bf n'}^*(\mathbf{r}'),
\end{equation*}
where $\rho_{\bf n, \bf n'}$ are the matrix elements of the one-particle reduced density matrix (OPRDM), with quantum number ${\bf n}\equiv(n,m)$ labeling the single-particle basis function specifying the radial and the angular momentum quantum numbers. The $\rho(\mathbf{r},\mathbf{r}')$ carries information about the traced out degrees of freedom of remaining ($N$-1) particles through $\rho_{\bf n, \bf n'}$. Since the OPRDM is Hermitian, it can be diagonalized to be written as

\begin{equation*}
\rho(\mathbf{r},\mathbf{r}')
=
\sum_\mu \lambda_\mu  \chi_\mu(\mathbf{r})\chi_\mu^*(\mathbf{r}'),
\end{equation*}
where $\{\chi_\mu(\mathbf{r})\}$ are the eigenvectors (called natural
orbitals \cite{Soumya2015}) of the OPRDM given by

\begin{equation*}
\chi_\mu(\mathbf{r})
=
\sum_ {\bf n} c_{\bf n}\,u_{\bf n}(\mathbf{r}),
\end{equation*}
with the corresponding eigenvalues satisfying the normalization condition

\begin{equation*}
\sum_\mu  \lambda_\mu = 1, \sum_\mu  \lambda_\mu m_{\mu} = L_z,
\quad
1\geq\lambda_1\cdots
\geq\lambda_\mu\geq\cdots\geq 0.
\end{equation*}
Each eigenvector $\chi_\mu(\mathbf{r})$ represents a fraction $\lambda_\mu$ of the condensate-noncondensate and $\chi_1(\mathbf{r})$ corresponding to the largest eigenvalue $\lambda_1$ is the Bose-Einstein condensate, equivalent to the macroscopic Gross-Pitaevskii wavefunction in the mean-field approximation \cite{dalfovo1999}. The vorticity of the Bose-Einstein condensate can be identified with the angular
momentum quantum number $m_1$.

\subsection{von Neumann entanglement entropy}
The von Neumann entanglement entropy provides a measure of non-local correlations between the constituting quantum particles \cite{Islam2015, eisert2010, Zhao2009, Zhao2010} which, over the years, has become an important tool for investigating and characterizing many-body quantum states.

To study the entanglement of subsystem \(A\) with subsystem \(B\) of the combined system $A+B$, one traces out the degrees of freedom of subsystem \(B\) to obtain the reduced density matrix of subsystem \(A\):
\begin{equation*}
    \hat \rho_A = \mathrm{Tr}_B(\hat \rho),
\end{equation*}
where $\hat \rho$ is the density matrix of the combined system $A+B$. The von Neumann entanglement entropy of subsystem $A$ is obtained as

\begin{equation}
    S^A_{ent} = -\mathrm{Tr}(\hat \rho_A \ln \hat \rho_A).
    \label{ent}
\end{equation}
The von Neumann entropy provides a measure of the entanglement of the subsystem $A$ with the rest of the system. In our work, $A$ is the one-particle subsystem and the remaining ($N$-1) particles form the subsystem $B$.

The lower value of entanglement entropy implies localization while the higher value corresponds to delocalization in the many-body Hilbert space \cite{Haibin2004, Lea2004, YAMAGUCHI2019}.

\section{Results and Discussion} \label{3}
We consider a system of $N = 4, 8, 12$ and $16$ bosonic atoms of $^{87}\mathrm{Rb}$ with radial frequency of the trap $\omega_{\perp} = 110 \pi$ \cite{talib2025, talib2025spectral, Dalfovo1996}. This choice of parameter and mass $M$ of the $^{87}\mathrm{Rb}$ atom yield radial trap length $a_{\perp} = \sqrt{\frac{\hbar}{M \omega_{\perp}}} = 1.446~\mu\text{m}$. The interaction range of the Gaussian potential is fixed at $\sigma = 0.1\,a_{\perp}$. The effective interaction strength in the mean-field approximation for contact potential is characterized by the parameter $\frac{N a_{sc}}{a_{\perp}}$ \cite{dalfovo1999}, where $a_{sc}$ is the s-wave scattering length. We employ exact diagonalization to obtain variationally exact lowest eigenvalues and the corresponding eigenvectors \cite{Ahsan2001} for a few-boson system. Owing to exponential increase of the many-body Hilbert space with increasing number of bosons $N$, our calculations are restricted to systems with a few tens of bosons. Thus, to achieve values of the parameter $\frac{N a_{sc}}{a_{\perp}}$ relevant to the experimental situations, we vary $a_{sc}$ to take values $a_{sc} = 10\,a_{0}, 100\,a_{0}, 1000\,a_{0}$ corresponding to the moderate interaction regime and $a_{sc}=10000\,a_{0}, 100000\,a_{0}$, $1000000\,a_{0}$ corresponding to the strong interaction regime, where $a_{0} = 0.05292~\text{nm}$ is the Bohr radius. Thus, the two-body interaction parameter defined in Eq. (\ref{g2}) takes values $g_{2} = 0.003669, 0.03669, 0.3669$ for moderate interaction regime, when the interaction energy is small compared to the trap energy, and $g_2=3.669, 36.69$, $366.9$ for strong interaction regime,  when the interaction energy is comparable to the trap energy.  

We now present our results on the structure of eigenstates using IPR, information entropy and von Neumann entanglement entropy.

\begin{table*}[t]
\caption{Values of the condensate fraction for $N=4, 8, 12$ and $16$ bosons for various values of the interaction strength $g_2$ in the non-rotating case.}
\label{CFtable}
\centering
\begin{ruledtabular}
\begin{tabular}{c c c c c c c}
$N$ &
\multicolumn{3}{c}{Moderate interaction} &
\multicolumn{3}{c}{Strong interaction} \\

& $g_2=0.003669$ & $g_2=0.03669$ &
$g_2=0.3669$ & $g_2=3.669$ & $g_2=36.69$ & $g_2=366.9$ \\

\cline{2-4} \cline{5-7}

4  & 0.99999990 & 0.99999041 & 0.99905425 & 0.91817960 & 0.23069971 & 0.04323045 \\
8  & 0.99999978 & 0.99997761 & 0.99776032 & 0.72078181 & 0.05262812 & 0.06036494 \\
12 & 0.99999965 & 0.99996479 & 0.99638549 & 0.44636006 & 0.04104913 & 0.10816060 \\
16 & 0.99999952 & 0.99995195 & 0.99487448 & 0.30610309 & 0.07779239 & 0.17720928 \\
\end{tabular}
\end{ruledtabular}
\end{table*}

\begin{figure}[t]
\centering
\includegraphics[width=1\linewidth]{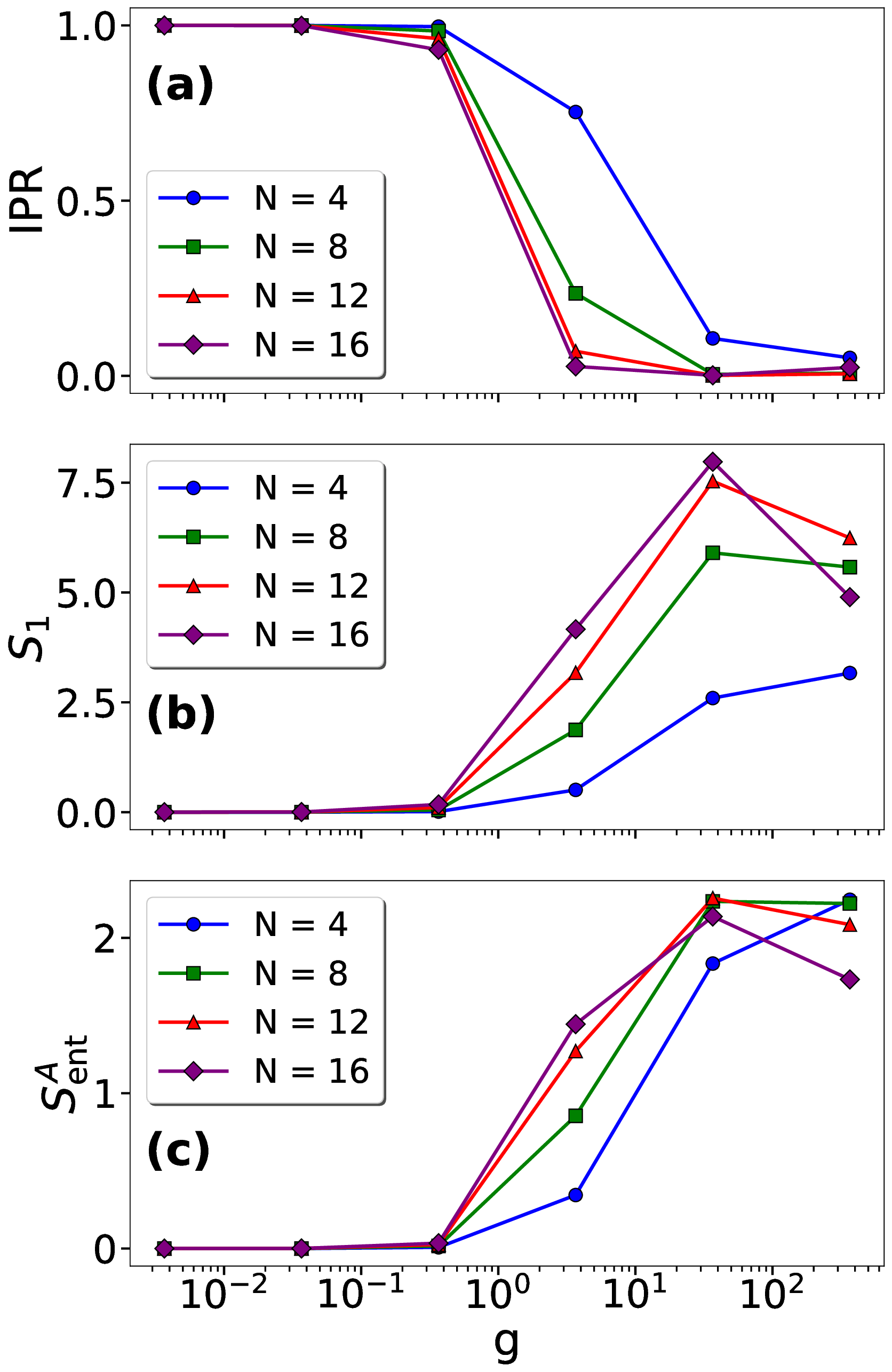}
\caption{Variation of (a) inverse participation ratio, (b) information entropy, and (c) von Neumann entanglement entropy, with interaction strength $g$ for $N=4, 8, 12, 16$ bosons in the non-rotating $L_z=0$ case.}
\label{fig:ps0}
\end{figure}

\subsection{Non-rotating case}

The inverse participation ratio (IPR) for different number of bosons in the non-rotating case is presented in Fig.~\ref{fig:ps0}(a). In the moderate interaction regime with $g_2 = 0.003669$, $0.03669$, and $0.3669$ the IPR remains close to $1$ for all considered systems with $N = 4, 8, 12,$ and $16$, indicating that the many-body ground state is strongly localized due to Bose-Einstein condensation, as indicated by the values of the condensate fraction given in Table. \ref{CFtable}.  This reflects the fact that the interaction strength is too weak to induce significant distribution of eigenstate weight over the many-body basis states. As the interaction strength is increased to $g_2 =3.669, 36.69$, $366.9$ in the strong interaction regime leading to increasing depletion of Bose-Einstein condensate, a systematic decrease in IPR values is observed, as shown in Fig. \ref{fig:ps0}(a) and Table. \ref{IPRtable}. The reduction in values of IPR implies that the eigenstate weight is spread over a significant number of many-body basis states, resulting in a delocalized eigenstate. 

In addition to the dependence on interaction strength, a system-size dependence is also observed. For a given value of the interaction strength, the IPR decreases as the number of bosons increases from $N = 4$ to $N = 8, 12,$ and $16$, as summarized in Table. \ref{IPRtable}. The trend indicates that the degree of delocalization becomes more pronounced with increasing system-size due to the competing effects of interaction and quantum-statistics, which can be understood as follows. There is an exponential increase of the Hilbert space dimensionality with increasing number of bosons, which allows the eigenstate weight to distribute over a larger space. Further, the interaction energy ($\propto \frac{Na_{sc}}{a_{\perp}}$) increases with $N$ causing the bosons to get distributed over different fractions ($\lambda_1, \lambda_2, \lambda_3...$) of the condensate-noncondensate, leading to depletion of the Bose-Einstein condensate,  corresponding to the fraction $\lambda_1$. As a result, even for moderate interaction strengths, larger systems tend to exhibit stronger delocalization. These results demonstrate a crossover from localization at moderate interaction to delocalization at strong interaction, with the transition becoming sharper as the number of bosons increases.
\begin{table*}[!t]
\caption{Values of the inverse participation ratio for $N=4, 8, 12$ and $16$ bosons for various values of the interaction strength $g_2$ in the non-rotating case.}
\label{IPRtable}
\centering
\begin{ruledtabular}
\begin{tabular}{c c c c c c c}
$N$ &
\multicolumn{3}{c}{Moderate interaction} &
\multicolumn{3}{c}{Strong interaction} \\

& $g_2=0.003669$ & $g_2=0.03669$ &
$g_2=0.3669$ & $g_2=3.669$ & $g_2=36.69$ & $g_2=366.9$ \\

\cline{2-4} \cline{5-7}

4  & 0.99999964 & 0.99996548 & 0.99660257 & 0.75290945 & 0.10676107 & 0.05121191 \\
8  & 0.99999840 & 0.99983882 & 0.98406810 & 0.23536448 & 0.00439641 & 0.00758336 \\
12 & 0.99999620 & 0.99961988 & 0.96214275 & 0.07013423 & 0.00117150 & 0.00578139 \\
16 & 0.99999308 & 0.99930840 & 0.93040020 & 0.02680354 & 0.00138771 & 0.02409025 \\
\end{tabular}
\end{ruledtabular}
\end{table*}

\begin{figure}[!t]
\centering
\includegraphics[width=1\linewidth]{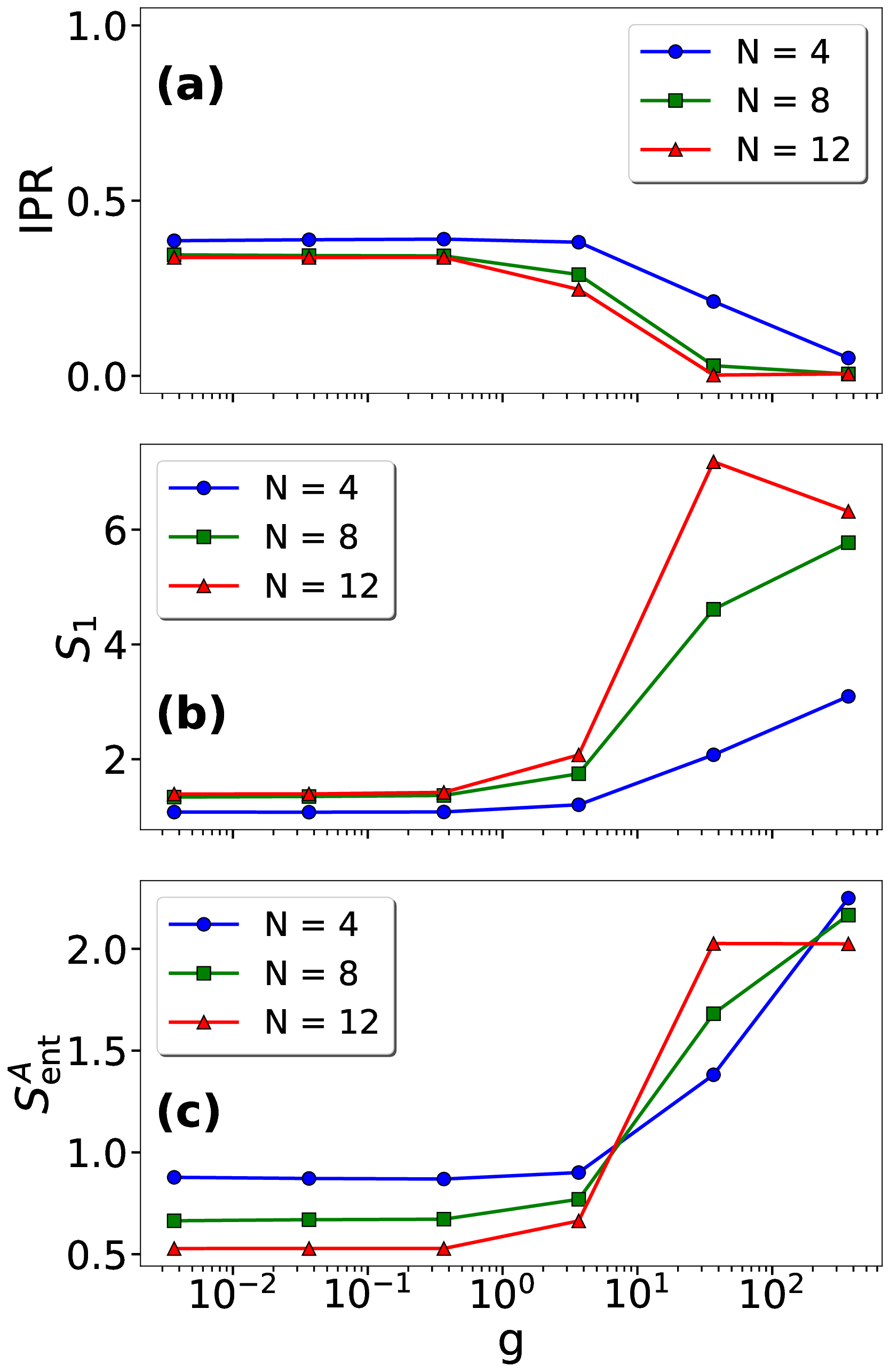}
\caption{Variation of (a) inverse participation ratio, (b) information entropy, and (c) von Neumann entanglement entropy, with interaction strength $g$ for $N=4, 8, 12$ bosons in the total angular momentum state $L_z=N$.}
\label{fig:ps2}
\end{figure}

\begin{figure}[!t]
\centering
\includegraphics[width=1\linewidth]{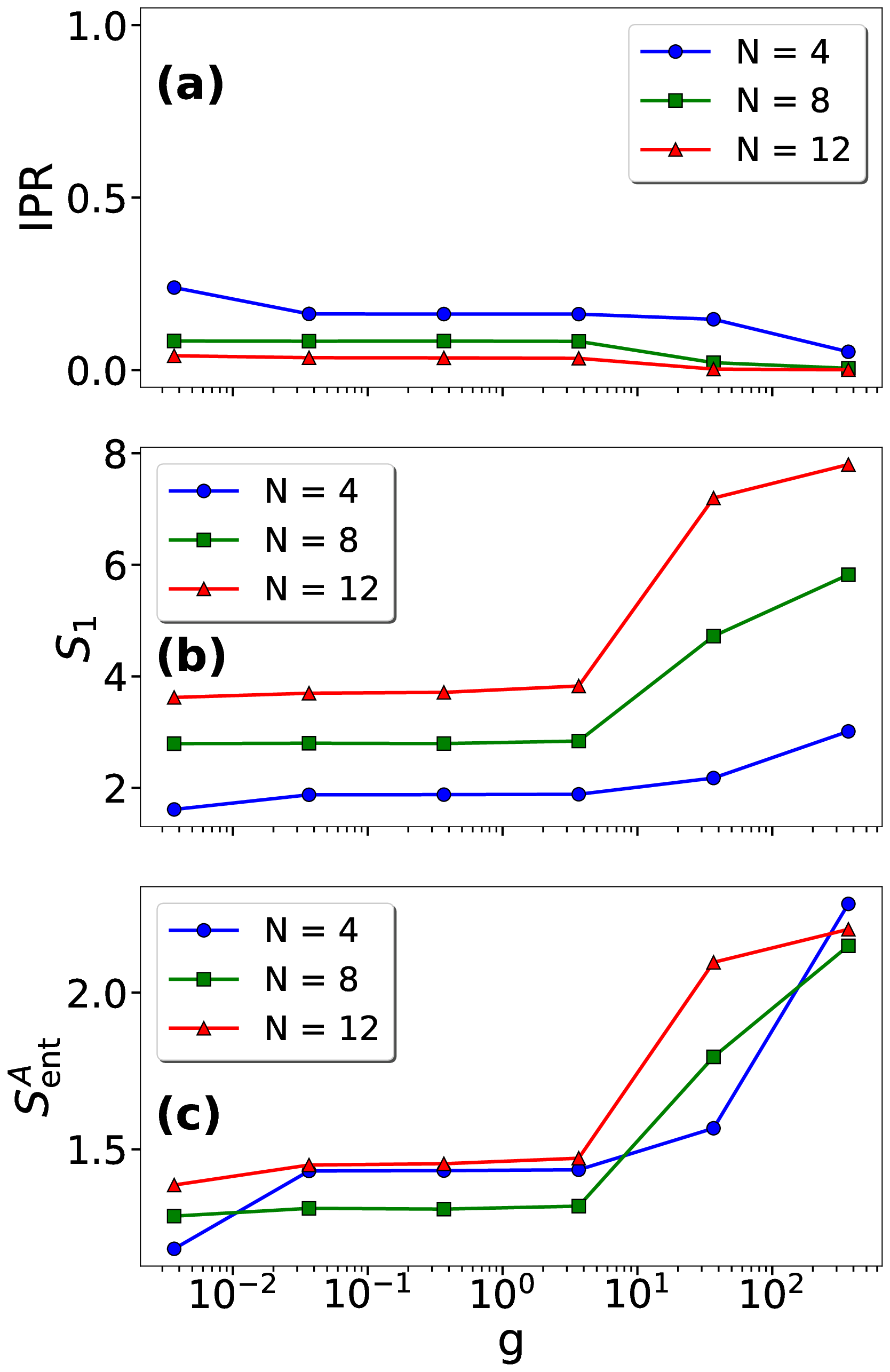}
\caption{Variation of (a) inverse participation ratio, (b) information entropy, and (c) von Neumann entanglement entropy, with interaction strength $g$ for $N=4, 8, 12$ bosons in the total angular momentum state $L_z=2N$.}
\label{fig:ps3}
\end{figure}

\begin{figure}[!t]
\centering
\includegraphics[width=1\linewidth]{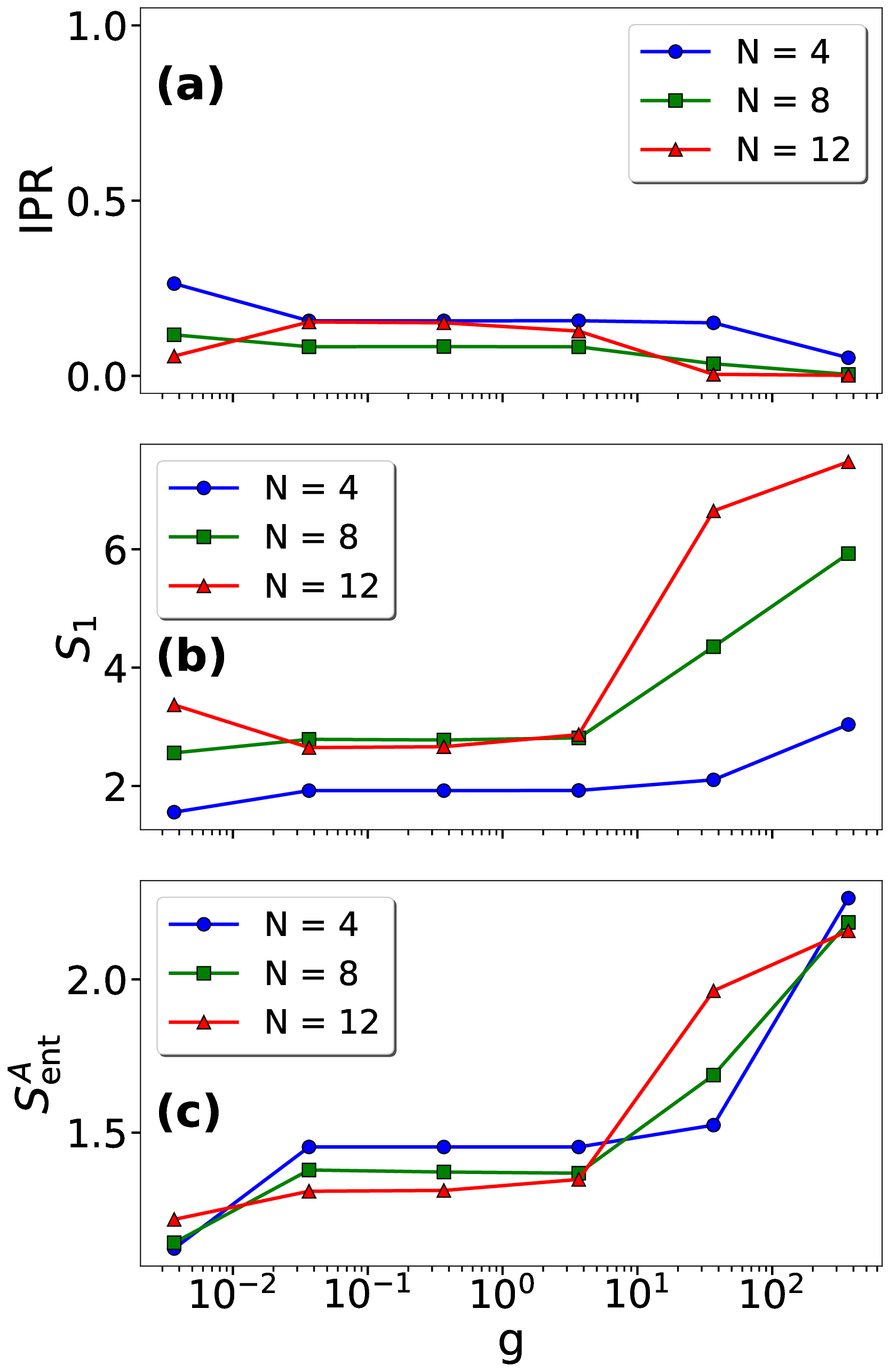}
\caption{Variation of (a) inverse participation ratio, (b) information entropy, and (c) von Neumann entanglement entropy, with interaction strength $g$ for $N=4, 8, 12$ bosons in the total angular momentum state $L_z=3N$.}
\label{fig:ps4}
\end{figure}

\begin{table*}[t]
\caption{Values of the condensate fraction for $N=4$, $8$, and $12$ bosons for various values of the interaction strength $g_2$ in the rotating case.}
\label{CFtablerot}
\centering
\begin{ruledtabular}
\begin{tabular}{c c c c c c c c}
$L_z$ & $N$ &
\multicolumn{3}{c}{Moderate interaction} &
\multicolumn{3}{c}{Strong interaction} \\

& & $g_2=0.003669$ & $g_2=0.03669$ &
$g_2=0.3669$ & $g_2=3.669$ &
$g_2=36.69$ & $g_2=366.9$ \\

\cline{1-2} \cline{3-5} \cline{6-8}

\multirow{3}{*}{N}
& $4$  & 0.64312047 & 0.64881842 & 0.65363774 & 0.67367789 & 0.55164975 & 0.04155878 \\
& $8$ & 0.77855547 & 0.77622364 & 0.77680701 & 0.76160817 & 0.48862321 & 0.15279385 \\
& $12$ & 0.83954036 & 0.83960395 & 0.84158981 & 0.81769982 & 0.28500468 & 0.15065113 \\

%\hline
\cline{1-2} \cline{3-5} \cline{6-8}

\multirow{3}{*}{2N}
& $4$  & 0.57773774 & 0.40245659 & 0.40169289 & 0.40199819 & 0.39643052 & 0.02778208 \\
& $8$ & 0.51614462 & 0.48946604 & 0.49127331 & 0.49247584 & 0.39977516 & 0.21310841 \\
& $12$ & 0.42835749 & 0.35604549 & 0.35315232 & 0.35361199 & 0.15278383 & 0.13236130 \\

\cline{1-2} \cline{3-5} \cline{6-8}

\multirow{3}{*}{3N}
& $4$  & 0.60876960 & 0.37884477 & 0.37897906 & 0.38115255 & 0.39505630 & 0.03354442 \\
& $8$ & 0.60562784 & 0.40891535 & 0.41280415 & 0.42938511 & 0.43837899 & 0.15121838 \\
& $12$ & 0.55918258 & 0.28508660 & 0.28315772 & 0.26189580 & 0.27084759 & 0.14181057 \\

\end{tabular}
\end{ruledtabular}
\end{table*}

The variation of the information (Shannon) entropy with increasing number of bosons is depicted in Fig.~\ref{fig:ps0}(b). In the moderate interaction regime, specifically for $g_2 = 0.003669, 0.03669, 0.3669$, the information entropy remains nearly constant for $N = 4, 8, 12,$ and $16$ bosons. The observed behavior suggests that the eigenstate structure does not change significantly with increase in number of bosons, highlighting the dominant role of Bose-Einstein condensation. As the interaction strength increases further in the strongly interacting regime to values $g_2 = 3.669, 36.69$, $366.9$, the information entropy increases systematically with the number of bosons. In particular, for a given interaction strength, the information entropy increases as the number of bosons increases from $N=4$ to $8, 12$, $16$, as shown in Fig. \ref{fig:ps0}(b). Thus, an increase in information entropy implies an increase in delocalization, resulting from increase in Hilbert space dimensionality with the number of bosons. For a given number of bosons, as the interaction strength increases, the Bose-Einstein condensate depletes progressively, as indicated by values of the condensate fraction shown in Table. \ref{CFtable}, thus resulting in delocalization of the eigenstate. The above results signify transition from a regime of moderately interacting, localized eigenstate with nearly constant entropy to a strongly interacting regime where the entropy increases with number of bosons, signaling the emergence of delocalized state.

The behavior of the von Neumann entanglement entropy exhibits trends that closely reflect the underlying localization-delocalization transition of the many-body system. In the moderate interaction regime, the von Neumann entropy, defined in Eq. (\ref{ent}), takes small values for $N = 4, 8, 12,$ and $16$ bosons, as shown in Fig. \ref{fig:ps0}(c), implying that bosons within the system are minimally entangled and the many-body eigenstate weights are distributed over only a few basis states, indicating many-body localization in the Hilbert space. As the interaction increases to strong regime, a significant enhancement in the von Neumann entropy is observed in Fig. \ref{fig:ps0}(c), implying an increasing entanglement between the bosons with interaction and crossover to many-body delocalization. For a given interaction strength in the strongly interacting regime, the von Neumann entropy increases systematically with the number of bosons in the system, leading to strong entanglement between the bosons implying progressive transition towards many-body delocalization. The observed behavior of the von Neumann entropy with number of bosons indicates  a transition from weakly entangled bosons in moderate interaction regime to highly entangled bosons in strong interaction regime.

\begin{table*}[!t]
\caption{Values of the inverse participation ratio for $N=4$, $8$, and $12$ bosons for various values of the interaction strength $g_2$ in the rotating case.}
\label{IPRtablerot}
\centering
\begin{ruledtabular}
\begin{tabular}{c c c c c c c c}
$L_z$ & $N$ &
\multicolumn{3}{c}{Moderate interaction} &
\multicolumn{3}{c}{Strong interaction} \\

& & $g_2=0.003669$ & $g_2=0.03669$ &
$g_2=0.3669$ & $g_2=3.669$ &
$g_2=36.69$ & $g_2=366.9$ \\

\cline{1-2} \cline{3-5} \cline{6-8}

\multirow{3}{*}{N}
& $4$  & 0.38555614 & 0.38836004 & 0.39007723 & 0.38128141 & 0.21212228 & 0.05091107 \\
& $8$  & 0.34531715 & 0.34322985 & 0.34229031 & 0.28887119 & 0.02889249 & 0.00529345 \\
& $12$ & 0.33785196 & 0.33782464 & 0.33805954 & 0.24658558 & 0.00220831 & 0.00556181 \\

%\hline
\cline{1-2} \cline{3-5} \cline{6-8}

\multirow{3}{*}{2N}
& $4$  & 0.23928943 & 0.16280831 & 0.16236975 & 0.16223672 & 0.14727138 & 0.05305504 \\
& $8$  & 0.08424884 & 0.08364455 & 0.08401686 & 0.08321494 & 0.02148003 & 0.00511589 \\
& $12$ & 0.04145137 & 0.03585959 & 0.03530396 & 0.03409133 & 0.00276863 & 0.00087965 \\

\cline{1-2} \cline{3-5} \cline{6-8}

\multirow{3}{*}{3N}
& $4$  & 0.26317315 & 0.15690733 & 0.15696335 & 0.15716815 & 0.15121746 & 0.05161524 \\
& $8$  & 0.11707692 & 0.08305780 & 0.08347803 & 0.08278011 & 0.03455206 & 0.00408900 \\
& $12$ & 0.05630556 & 0.15372592 & 0.15131035 & 0.12758776 & 0.00415871 & 0.00143792 \\

\end{tabular}
\end{ruledtabular}
\end{table*}

\subsection{Rotating case with various angular momenta}

The variation of the inverse participation ratio (IPR) in the moderate interaction regime for $N = 4, 8, 12$ rotating bosons with total angular momentum $L_z = N$, is shown in Fig.~\ref{fig:ps2}(a). The IPR decreases from values close to unity for the non-rotating case to less than half the value for the single-vortex state $L_z = N$, indicating transition from relatively localized state to delocalized state due to depletion of the Bose-Einstein condensate. The values of the condensate fraction are given in Table. \ref{CFtablerot}. The IPR remains nearly constant in the moderate interaction regime and decreases in the strong interaction regime, as seen from Fig.~\ref{fig:ps2}(a) and summarized in Table. \ref{IPRtablerot}. The behavior suggests that, in the single-vortex state $L_z=N$, the system exhibits a relatively stable degree of delocalization in the moderate interaction regime, followed by increasing distribution of the eigenstate weight over the Hilbert space in the strong interaction regime. This increased delocalization can be attributed to the depletion of the Bose-Einstein condensate resulting from strong interaction as well as nucleation of the single-vortex state, as seen from the condensate fraction summarized in Table. \ref{CFtablerot}. For higher angular momentum states, $L_z = 2N$ and $3N$, the IPR values are consistently lower than those observed for $L_z = N$, as shown in Figs. \ref{fig:ps3}(a) and \ref{fig:ps4}(a), indicating a higher degree of delocalization induced by rotation resulting in nucleation of multi-vortices \cite{hamid2022}. 

The information entropy remains largely unchanged in the moderate interaction regime and rises steeply in the strong interaction regime indicating increasing degree of delocalization, for $N = 4, 8, 12$ bosons in the single-vortex state $L_z = N$, as shown in Fig. \ref{fig:ps2}(b). Further, for higher angular momentum states, $L_z = 2N$ and $3N$, the information entropy attains larger values compared to $L_z = N$ state, as shown in Figs. \ref{fig:ps3}(b) and \ref{fig:ps4}(b),  indicating a greater degree of delocalization induced by rotation. 

To examine the entanglement between the bosons in the system, we present the von Neumann entropy for $N = 4, 8, 12$ in the total angular momentum single-vortex state $L_z = N$, as shown in Fig. \ref{fig:ps2}(c). The von Neumann entropy remains nearly constant and takes small values in the moderate interaction regime, indicating that the bosons are weakly entangled and less localized compared to the non-rotating case. As the interaction increases further to the strong interaction regime, the von Neumann entropy increases sharply, reflecting development of strong quantum correlations leading to enhanced entanglement between the bosons with the many-body eigenstate evolving towards delocalization. 
With increasing rotation to $L_z = 2N$ and $3N$ states, von Neumann entropy attains higher values compared to the single-vortex state $L_z = N$, as shown in Figs. \ref{fig:ps3}(c) and \ref{fig:ps4}(c),  demonstrating that rotation, leading to nucleation of multi-vortices, enhances the degree of entanglement within the system and consequent emergence of many-body delocalization in the Hilbert space.

As the number of bosons increases from $N = 4, 8$ to $12$, a systematic variation is observed in all the considered measures, namely the IPR, the information entropy and the von Neumann entanglement entropy, for all three values of the total angular momentum $L_z$, as shown in Figs. \ref{fig:ps2}, \ref{fig:ps3} and \ref{fig:ps4}. In particular, the IPR decreases with number of bosons, as summarized in Table. \ref{IPRtablerot}, indicating enhanced delocalization of the eigenstate, while the corresponding entropy measures reflect an overall change in the structure and correlations of the many-body eigenstate. 

\section{Summary and Conclusion}\label{4}

In the present work, for the many-body ground state,  we have investigated localization-delocalization crossover and entanglement within a system of interacting bosons in the non-rotating and rotating cases, by analyzing the inverse participation ratio, information (Shannon) entropy and the von Neumann entanglement entropy.

For the non-rotating case, in the moderate interaction regime, the IPR takes values close to unity while the information entropy and the von Neumann entanglement entropy take small and nearly constant values, indicating that the ground state is localized in the many-body Hilbert space, with bosons remaining weakly entangled, due to Bose-Einstein condensation. As the interaction increases to strong regime, the IPR decreases and the information and the entanglement entropies increase steeply, reflecting progressive distribution of the eigenstate weight over the Hilbert space and the emergence of enhanced entanglement, arising from the depletion of Bose-Einstein condensate. The behavior becomes more pronounced with increasing number of bosons, due to exponential increase of dimensionality of the many-body Hilbert space, demonstrating enhancement in delocalization and consequently strong entanglement within the system. Thus, the system exhibits a crossover from many-body localization to delocalized phase as the interaction strength increases from moderate to strong regime, with the effect being further amplified by increasing number of bosons.

For the rotating case, the single-vortex state $L_z=N$ exhibits decrease in IPR and increase in information and von Neumann entropies compared to the corresponding values in the non-rotating case, resulting in enhanced delocalization with increasing interaction, arising from further depletion of the Bose-Einstein condensate due to formation of the single-vortex state. For higher angular momentum states with $L_z=2N$ and $L_z=3N$, resulting in nucleation of multiple vortices, the IPR remains consistently lower while the information and the von Neumann entropies take higher values with respect to the values for the single-vortex state, indicating an increasing degree of delocalization and enhancement of entanglement with increasing interaction, arising from further depletion of Bose-Einstein condensate due to nucleation of multi-vortices.

Thus, all the three quantities considered in the present work---IPR, information entropy and von Neumann entropy---exhibit consistent qualitative behavior across the moderate and the strong interaction regimes. The close agreement between these measures highlights their common sensitivity to the underlying localization-delocalization crossover. In particular, the information entropy indicates an approach towards delocalized behavior in the strong interaction regime, while increase of von Neumann entropy signals the development of significant quantum entanglement within the system. The present study, thus, provides a unified picture of how interaction, rotation, and number of bosons govern the crossover from localization to delocalization. 

The system of trapped interacting bosons under external rotation investigated in the present work offers a viable platform where many-body angular momentum states could be employed as metrological probes for rotation sensing \cite{dengis2026, baak2026}.

\section*{Acknowledgments}M.T. thanks the Ministry of Social Justice and Empowerment, Government of India, for Senior Research Fellowship (SRF).

\nocite{*}

\bibliography{apssamp}

\end{document}